# The development of food protein-inorganic hybrid nanoflowers with outstanding role in stabilizing natural pigments

Penghui Shen[1, 2, 3], Mouming Zhao[1, 2], Jasper Landman[3], Feibai Zhou[1, 2]*

[1] School of Food Science and Engineering, South China University of Technology, Guangzhou 510640, China

[2] Guangdong Food Green Processing and Nutrition Regulation Technology Research Center, Guangzhou 510640, China

[3] Laboratory of Physics and Physical Chemistry of Foods, Wageningen University, Bornse Weilanden 9, 6708 WG Wageningen, The Netherlands

* Corresponding author

Email address: penghui.shen@wur.nl (Penghui Shen), feibaizhou@scut.edu.cn (Feibai Zhou), jasper.landman@wur.nl (Jasper Landman).

# Abstract

Protein-inorganic hybrid nanoflowers (HNFs) possess unique properties in promoting surface reaction and have attracted wide-spread attention as a newly developed nanomaterial. However, the availability of protein sources has up to now been mostly limited to enzymes, which narrows the application of HNFs especially in food industry. Here we show that for many types of food protein, enzymatic hydrolysis can improve its ability to form versatile HNFs, or even induce HNF formation where the protein source did not show its formation *a priori*. The treatment of enzymatic hydrolysis increases the flexibility of such proteins and induces nucleation sites of HNFs in the early formation stage by decomposing those proteins into polypeptides. In particular, the HNF prepared with soy protein hydrolysate further shows a high loading capacity of water-soluble *Monascus* red, reaching up to 554.1 mg per gram of HNF. Its stabilization towards lipophilic curcumin is similarly impressive with the loading capacity reaching 21.9 mg per gram of HNF. This HNF could also effectively protect these two sensitive natural pigments in harsh environments. This research significantly broadens the available protein source for HNF fabrication and demonstrates the potential of HNFs as novel protein-based delivery systems – especially for sensitive natural pigments – which could serve the fields of food, cosmetic and medicine.

# Introduction

As a newly developed aesthetic nanomaterial, nanoflowers are rapidly blooming in the scientific world due to their distinctive property of high surface-to-weight ratio, which markedly improves the efficiency of surface reactions[1]. Early nanoflowers were mostly prepared directly from inorganic materials, such as carbon[2], elemental metals[3,4], metal alloys[5] and metal compounds[6,7], and were primarily applied as optoelectronics devices or sensors in catalysis and solar cells[8]. However, their preparation methods generally involve complex treatments that need various organic solvents, high temperature and high pressure[8].

During last decade, a novel hybrid nanoflower (HNF) was accidently discovered in mixtures of $CuSO_4$ solution and phosphate buffer containing bovine serum albumin (BSA)[9]. Here, BSA protein coordinates to the generated insoluble copper phosphate and easily forms the hierarchical HNF after leaving the mixture standing. This means that preparation of nanoflowers could be much easier and safer. This discovery also promoted the development of organic-inorganic hybrid nanoflowers[1,10-13], as novel nanomaterials showing great promise in the fields of biomedical, biosensors, biocatalysis, bioenergy and environment. Currently, new strategies mostly focus on the immobilization of enzymes, which could simultaneously improve their stability and activity[10,12,14].

In recent years, researchers have attempted to develop HNFs as delivery systems. Polysaccharides[15] and polyphenols[16] were explored as the organic materials to carry some small biomolecules (e.g. 5-Flurouracil and curcumin). Protein is one of the most used biomaterials in the design of delivery systems because it has abundant functional groups that can combine with other biomolecules through hydrophobic interactions, hydrogen bonds and electrostatic interactions[17-19]. Once incorporated into high surface-to-weight ratio HNF, protein could contribute to a better stabilization of loaded cargos due to the synergistic effects of immobilized proteins as shown in aforementioned enzyme-based HNFs[20-23]. Compared to enzymes, food proteins show great potential for fabricating delivery systems because they are more abundant and cost-effective. However, few food proteins have been shown to effectively produce HNFs, and are limited to BSA[24,25], concanavalin A[26], immunoglobulin G[27], human serum albumin[28], hemoglobin[29], and streptavidin[30]. Moreover, in most cases these proteins are used because of their special enzyme mimicking activities or specific affinity with some substrates (especially antibodies), and the prepared HNFs are mainly applied as biomimetic catalysts or biosensors. We hypothesize that the limited use of food proteins for HNF fabrication is caused by the high molecule weight and less flexible structures of general food proteins (particularly for plant proteins) compared to enzymes.

In this article, we investigate 11 proteins from common food materials including pulses (pea and mung bean), oil seeds (soy bean and peanut), cereals (oat, rice and brown rice), root (potato), algae (chlorella) and milk (whey protein and casein). We demonstrate that these proteins indeed have poor ability to form HNFs. However, decomposition of these food proteins by enzymatic hydrolysis can markedly promote the formation of highly hierarchical HNFs with different morphologies. We show that soy protein hydrolysate (SPH) can combine with various metal ions and form different types of HNF with sizes spanning between 25 nm and 40 μm. These results indicate that enzymatic hydrolysis can act as a powerful tool in the development and design of food protein-inorganic hybrid nanoflowers. In addition, we show that SPH-based HNF can be loaded with a large quantity of pigments: per gram, hybrid nanoflowers can carry 554.1 mg *Monascus* red or 21.9 mg curcumin. Moreover, this HNF also showed excellent performance in stabilizing these natural pigments against UV light and high temperatures. This study has the potential to facilitate the application of a wide range of ordinary but abundant food proteins in the fabrication of new protein-inorganic hybrid nanoflowers. In addition, this study provides a new and effective strategy for the stabilization of natural small biomolecules in novel delivery systems that could be further applied in food, cosmetic and medicine.

## Formation of hybrid nanoflowers by soy protein hydrolysate

We firstly explore the effect of enzymatic hydrolysis on soy protein isolate (SPI) to form hybrid nanoflowers (HNFs) as shown in Fig. 1a. We see that before enzymatic hydrolysis, HNFs barely form from the primary precipitates formed by SPI and copper phosphate (a1), although in our experiments most (61.6% ± 0.7%, w/w) of the SPI had been immobilized, indicating the inferior performance of SPI in fabricating HNF. After enzymatic hydrolysis, HNFs started to form (a2-a4). The number of these HNFs evidently increased with the enzyme/substrate (E/S) ratio increasing, particularly at the ratio of 0.15% (a4), where the primary precipitates were mostly converted into HNFs. These HNFs display homogeneous round appearance and had diameter around 10 μm. Meanwhile, 68.8% (± 2.2%, w/w) of soy protein hydrolysate (SPH) was immobilized at the E/S ratio of 0.15%, only slightly higher than with SPI, suggesting enzymatic hydrolysis could effectively improve the capability of SPI in producing HNFs. We recognize that the hydrolysis enzyme may participate in the formation of HNFs, but given the overall low total enzyme content of no more than 0.15% of the total weight of soy protein, we estimate its effect to be rather small.

Structural characterization of soy protein hydrolysates was further evaluated for better understanding the formation of HNF. SDS-PAGE under non-reducing conditions shows that as the E/S ratio increases, the protein components with high molecular weight at the top of stacking gel and running gel evidently diminishes, generating protein continuums with lower molecular weight (Fig. 1b). At the E/S ratio of 0.15%, the bands of native SPI are barely visible, indicating that SPI was highly degraded into polypeptides or small peptides. However, the degree of hydrolysis (DH) of SPI is overall low and only reached 2.77% (± 0.09%) at the E/S ratio of 0.15%, suggesting the structure of SPI is hydrolyzed to mostly large polypeptides, and only a negligible fraction of shorter peptides or free amino acids are generated. This is attributed to the high efficiency of papain in hydrolyzing SPI, and similar phenomena were also reported by previous studies[31,32]. In addition, the hydrolysates and SPI share similar secondary structures (Supplementary Table 1) as shown by the Fourier-transform infrared spectroscopy (FTIR) spectra at the amide I region (1600-1700 cm-1) (Fig. 1d). Moreover, the intrinsic emission fluorescence spectra of SPI only show a small red shift (2.5 nm) and only a 6% decrease of the maximum intensity with the E/S ratio increasing to 0.15% (Fig. 1e; Supplementary Table 2). These results indicate that after the hydrolysis, the microenvironment of the tryptophan in SPI becomes slightly more hydrophilic, and the tertiary structure of SPI is only slightly altered. The above findings demonstrate that SPI is easily and simply decomposed into polypeptides by papain especially at an E/S ratio of 0.15%, where only the quaternary structure of SPI is largely altered. These polypeptides could have more binding sites towards copper phosphate compared to SPI, thus they might act as efficient building blocks facilitating the formation of HNFs. Additionally, further increase in hydrolysis degree caused hydrophobic aggregation and the resultant SPHs were therefore no longer considered.

The growth process of HNFs prepared with SPH at E/S ratio of 0.15% (Supplementary Fig. 1) was further recorded for better understanding the formation of HNFs. Upon addition of SPH to phosphate buffer and CuSO$_4$ solution, primary copper phosphate precipitates are generated first (a, f), which in turn transform into agglomerates within 3 h (b, g), probably facilitated by the coordinating ability of the amide groups in the protein backbone to $Cu^{2+}$[9,16]. Meanwhile, the agglomerates can provide locations for nucleation, initiating the growth of HNFs, where scaffolds appear in branched structures. At 12 h, a few HNFs finished their growth, with the scaffold structures becoming plump since proteins serve as the 'flesh' to fill the scaffolds (c, h). After 1d, many more HNFs have become visible (d, i), and at 3d, all precipitates seem to have converted into HNFs (e, j), indicating the complete generation of HNFs, in line with the formation period of general organic-inorganic HNFs[12,33]. Overall, the growth processes of SPH-copper HNF are similar to those of other organic-inorganic HNFs as previously reported. In the following sections, the SPH obtained at E/S ratio of 0.15% and the formation time of 3d were chosen to further modulate the generation of HNF.

## HNFs formation capability with varied metal ions

We further explore the potential of SPH to fabricate HNF in combination with different metal ions including $Mn^{2+}$, $Fe^{2+}$, $Ca^{2+}$, $Zn^{2+}$ and $Fe^{3+}$ under 0.8 mM and 2.5 mM concentrations, respectively. Indeed, versatile organic-inorganic HNFs have been successfully fabricated in the past by altering the metal ion species[34-36]. Interestingly, the macroscopic views (insets of Figure 2a, b) show that the addition of $Cu^{2+}$, $Mn^{2+}$ or $Fe^{2+}$ generates organized precipitates at both metal ion concentrations (insets of a1-a3, b1-b3), while the addition of $Ca^{2+}$ only generates discernable organized precipitates at the higher concentration (insets of a4 and b4). Herein, precipitates formed with $Cu^{2+}$ are coarse and granular (insets of a1, b1), while those formed upon $Mn^{2+}$ are integrated and lamellar (insets of a2, b2). Addition of $Zn^{2+}$ or $Fe^{3+}$ only produces loose and unorganized precipitates (insets of Supplementary Fig. 2a, b) and the influence of $Fe^{2+}$ and $Ca^{2+}$ is dosage-dependent (insets of a3 and b3, a4 and b4).

Under microscopic views, those organized precipitates are comprised of hierarchical HNFs (Figure 2a-d), while those unorganized precipitates are only comprised of loose and amorphous structures (Supplementary Fig. 2a, b). Specifically, HNFs prepared by 0.8 mM $Cu^{2+}$ display round and porous structures with diameter around 12 μm under field-emission scanning electron microscope (FE-SEM) (Fig. 2c1), in line with the result from optical light microscope (OLM) (Fig. 2a1). At 2.5 mM $Cu^{2+}$, the morphology of HNFs becomes a full-blossom state with large petals, and the average size of the HNFs increases to 26 μm (Fig. 2b1, d1). SPI also formed organized precipitates and HNFs with 2.5 mM $Cu^{2+}$ (Supplementary Fig. 3), but these HNFs are larger, less homogeneous and less regular compared to the HNFs prepared with SPH (Fig. 2b1, d1). The HNFs formed at 0.8 mM MnCl$_2$ are quite uniform and spherical with small and staggered petal-like structures, and the diameter is around 9 μm (Fig. 2a2, c2). While at 2.5 mM $Mn^{2+}$, the morphology of HNFs distinctly changes to donut-like structures, and the diameter decreases to around 5 μm (Fig. 2b2, d2). The HNFs prepared by $Fe^{2+}$ consist of bud-like agglomerates at both metal ion concentrations, and the bud size increases from around 25 nm to around 45 nm when the concentration of $Fe^{2+}$ increases from 0.8 mM to 2.5 mM (Fig. 2a3-d3). As for $Ca^{2+}$, HNFs display flexible petal-like structures with diameter around 1 μm at the high concentration of $Ca^{2+}$ (Fig. 2d4). Overall, SPH can combine with various metal ions and form versatile HNFs. Both the metal ion species and metal ion concentrations determine the generation of HNFs with SPH and the macroscopic and microscopic morphologies of the HNFs.

Unexpectedly, we found that the anion ion species also influenced the fabrication of HNF, which is reported for the first time to the best of our knowledge. The HNF prepared with $CuCl_2$ displayed almost the same shapes but larger diameters at both metal ion concentrations (Fig. 2a5-d5) compared to those prepared with $CuSO_4$ (Fig. 2a1-d1). The diameter of $CuCl_2$-HNF was around 14 μm at 0.8 mM $CuCl_2$ (Fig. 2a5, c5) and around 40 μm at 2.5 mM $CuCl_2$ (Fig. 2b5, d5). This suggests that anion ion may also join in or adjust the self-assembling of HNF probably through the interaction with the cation metal ions[37].

Furthermore, we explore the role of SPH in constructing those different HNFs. Without the addition of SPH, the precipitates tend to be less granular or become fragmentary from original integrated states (insets of Supplementary Fig. 4), and the generated HNFs become less plump (Supplementary Fig. 4). The HNFs prepared with $CuSO_4$ or $CuCl_2$ display similar radial and needle-like structures at lower metal ion concentration (Supplementary Fig. 4a, e), and display branched and loose structures with increased diameters at the higher metal ion concentration (Supplementary Fig. 4f. j). The HNFs prepared with $Mn^{2+}$ form donut-like structures at both metal ion concentrations (Supplementary Fig. 4b, g); while at higher metal ion concentration, the structure becomes smaller and weaker (Supplementary Fig. 4g). This situation also happens on the HNFs prepared with $Fe_{2+}$ and $Ca^{2+}$ (Supplementary Fig. 4c, d, h, i). These results strongly demonstrate that SPH acts as "flesh" in constructing HNFs and make them more plump and solid, while the metal phosphate mainly forms the scaffolds of HNFs.

Element maps show that the elements of nitrogen (N), metal (Cu/Mn/Fe/Ca), oxygen (O) and phosphorus (P) are uniformly distributed throughout all HNFs at both metal ion concentrations (Fig. 3a). This indicates that SPH is homogeneously hybridized with varied metal ions, and the elements of O and P are preliminarily assigned to phosphate groups. Meanwhile, the signals of metal, O and P elements are distinctly denser and brighter than that of N element, concurring with the scaffold roles of the metal phosphates in HNFs. We further selected HNFs formed at lower metal ion concentration and that only formed at higher $Ca^{2+}$ concentration as the representatives for FTIR and X-ray diffraction (XRD) analysis. The FTIR spectra of all HNFs display similar characteristic absorption peaks with that of SPH at (in cm$^{-1}$) 2800-3000 (-$CH_2$ and -$CH_3$ stretching), 1645 (amide I, C=O stretching), 1543 (amide II, C-N stretching and N-H bending), 1448 ($CH_2$ bending), 1398 ($CH_2$ wagging), 1240 (amide III, C-N stretching and N-H bending) (Fig. 3c), firmly indicating the presence of SPH in HNFs. These FTIR spectra also display the characteristic absorption peaks of $PO_4^{3-}$ at (in cm$^{-1}$) 900-1200 and 530-670, confirming the existence of phosphate groups in HNFs. In XRD patterns, the characteristic diffraction peaks of HNFs prepared with $Cu^{2+}$, $Mn^{2+}$ or $Ca^{2+}$ match well with those in the corresponding standard JCPDS cards of $Cu_3(PO_4)_2 \cdot 3H_2O$ (No. 22-0548), $Mn_3(PO_4)_2 \cdot 3H_2O$ (No. 03-0426) and $Ca_3(PO_4)_2 \cdot nH_2O$ (No. 18-0303) (Fig. 3d1-d3, d5). As for the HNF prepared with $FeSO_4$, its XRD pattern displays decreased main diffraction peaks compared to the standard JCPDS card of $FePO_4$ (No. 29-0715) (Fig. 3d4), but both patterns are roughly matched. These results finally confirm the successful hybridization of SPH and varied metal ions.

## Natural pigments loading and stabilization

We selected HNF prepared at 0.8 mM $CuSO_4$ ($CuSO_4$-HNF) to further load and stabilize natural pigments for two reasons: first, this condition is the most commonly used condition for fabricating HNFs in numerous previous studies; second, HNF prepared at 0.8 mM $CuSO_4$ in this study also displays a high surface-to-weight ratio of around 56 m$^2$/g, distinctly higher than that prepared at 2.5 mM $CuSO_4$ (around 32 m$^2$/g). We chose two typical natural pigments—water-soluble *Monascus* red and oil-soluble curcumin—as representative cargos.

Surprisingly, one gram of $CuSO_4$-HNF loaded 554.1 (± 85.4) mg of *Monascus* red, indicating high affinity of the HNF with this natural water-soluble pigment. Curcumin could be also loaded in high quantities, up to 21.9 (± 1.9) mg per gram of $CuSO_4$-HNF. After being mixed with water (containing 0.03% xanthan gum to suspend $CuSO_4$-HNF; same for below) and yielding a net curcumin concentration of 6 μg/mL, the loaded curcumin remained well dispersed (Fig. 4a2), demonstrating a pronounced improvement of the dispersion of curcumin in water compared to its poor solubility of unbound curcumin in water – between 11-390 ng/mL[38,39]. Moreover, the loaded curcumin suspension almost kept the same color as free curcumin in a similarly concentrated aqueous solution (containing 1.8% ethanol at volume ratio to solubilize the pure curcumin) (Fig. 4a1). In contrast, the loaded *Monascus* red slightly became pink in water due to the interference of the blue color of $CuSO_4$-HNF (Fig. 4a6) compared to free *Monascus* red (Fig. 4a5). Meanwhile, two pigment suspensions became opaque (Fig. 4a2, a6) since micron-scale $CuSO_4$-HNF (Fig. 2a1, c1) scattered the light. Additionally, nearly all pigments precipitated after centrifugation (Fig. 4a4, a8), indicating the successful loading of these two types of natural pigments.

We further investigated the mechanisms by which the $CuSO_4$-HNF embeds these pigments. From FTIR spectra of the pigment-HNF complex, we observed the characteristic peaks from both pigments, such as those at (in cm$^{-1}$) 1516 and 1284 for curcumin-HNF (Supplementary Fig. 5a) and those at (in cm$^{-1}$) 1541, 1464, 1406 and 1205 for *Monascus* red-HNF (Supplementary Fig. 5b), confirming the existence of these pigments in the HNF. In the FTIR spectra of curcumin-HNF, many absorption peaks relevant to the hydrophobic benzene ring of curcumin disappeared after loading, such as those in the functional group region at (in cm$^{-1}$) 1628, 1603 and 1429, and those in the fingerprint region at (in cm$^{-1}$) 1284, 1205, 1155, 856 and 808[40]. Moreover, the micro environment of curcumin became much more hydrophobic as shown by the drastic blue shift of its intrinsic emission fluorescence spectrum after loading onto $CuSO_4$-HNF (Supplementary Fig. 6a). These results indicate that the driving force for the curcumin complexation is mainly hydrophobic interaction. In addition, the disappearance of the peaks at 3508 cm$^{-1}$ (-OH stretching) and 962 cm$^{-1}$ (-COH bending) in the FTIR spectrum (Supplementary Fig. 5a) implies the interaction of the hydroxyl group of curcumin with the $CuSO_4$-HNF, most likely through hydrogen bonds. As for *Monascus* red, its FTIR spectrum shows a distinctly reduced and shifted negative trough at around 3340 cm$^{-1}$ (–NH stretching), and the absorption peaks at 1080 cm$^{-1}$ (C-N bending) and 1026 cm$^{-1}$ (C-O bending) disappeared after complexation (Supplementary Fig. 5b). This suggests that *Monascus* red combines with $CuSO_4$-HNF mainly through hydrogen bonding. On the other hand, the FTIR spectrum of loaded *Monascus* red also exhibites shifted and/or diminished hydrophobic peaks at (in cm$^{-1}$) 2962 (-CH$_3$ stretching), 2929 (-CH$_2$ stretching), 2873 (-CH$_3$ stretching) and 1464 (-CH$_2$ stretching), and a blue shift occurred in the intrinsic emission fluorescence spectrum of loaded *Monascus* red (Supplementary Fig. 6b). These results reveal that hydrophobic interactions also exist between *Monascus* red and the $CuSO_4$-HNF.

Afterwards, we evaluated the stability of curcumin and *Monascus* red after being loaded by $CuSO_4$-HNF under the treatment of UV light or heating. Free curcumin degrades quite rapidly in both harsh environments (Fig. 4b1,c1), where only 14.13% (± 0.23%) of curcumin was left after 2 h of UV light treatment (Fig. 4b3), and no curcumin was retained after 2 h of heating (Fig. 4c3) due to its rather poor thermal stability in water [41,42]. On the contrary, the loaded curcumin was much more stable with its color mostly retained (Fig. 4b2, c2). We observe that up to 84.1% (± 0.3%) of curcumin is left intact after 2 h of UV light treatment (Fig. 4b3), and similarly, 41.4% (± 2.6%) of curcumin remained after 2 h of heating (Fig. 4c3). $CuSO_4$-HNF also protected *Monascus* red during UV light treatment to a remarkable degree, where the retention rate of *Monascus* red was enhanced to 79.7% (± 0.8%) (Fig. 4b6) compared to 37.6% (± 3.6%) of free *Monascus* red. During heating, loaded *Monascus* red seems to have similar stability compared to free *Monascus* red, where around 70% of *Monascus* red was retained for both cases. Actually, we observed that *Monascus* red could form acid-insoluble precipitates with $CuSO_4$-HNF during heating, which could not be extracted for detection. The amount increased with increasing heating time (Supplementary Fig. 7b), suggesting that $CuSO_4$-HNF might strongly interact with *Monascus* red during the heating, possibly through covalent interaction[43], which further protecting this fraction of *Monascus* red. Therefore, the actual retention rate of *Monascus* red in $CuSO_4$-HNF should be higher than free *Monascus* red, implying $CuSO_4$-HNF also improved the thermal stability of *Monascus* red. However, such a phenomenon did not happen on loaded curcumin. (Supplementary Fig. 7a)

During heating, we also observed a high release of *Monascus* red from $CuSO_4$-HNF to aqueous phase (Supplementary Fig. 8b). After 10 min of heating, the retention rate of *Monascus* red in $CuSO_4$-HNF largely decreased from 88.8% (± 1.7%) to 55.8% (± 1.9%), while the *Monascus* red content in aqueous phase largely increased from 11.2% (± 0.8%) to 38.9% (± 1.9%). The high temperature-dependence release behavior of *Monascus* red indicates the predominant role of hydrogen bonds in maintaining the combination of *Monascus* red with $CuSO_4$-HNF. This phenomenon aligns with the results from FTIR analysis (Supplementary Fig. 5b). Under UV light treatment, the release of *Monascus* red pigment from $CuSO_4$-HNF was not discernible (Supplementary Fig. 9b). As for curcumin, only a small amount released from $CuSO_4$-HNF during the heating (Supplementary Fig. 8a), and this did not happen in UV light treatment as well (Supplementary Fig. 9a), suggesting the hydrogen bonds play a small role in the complexation of curcumin and $CuSO_4$-HNF. Overall, $CuSO_4$-HNF owns good capability to load the natural pigments and could effectively protect them in the UV light or heating environment.

## Universality of enzymatic hydrolysis in promoting food proteins to form hybrid nanoflowers

The above results clearly demonstrate that the decomposition of SPI by enzymatic hydrolysis can effectively enhance its ability to form HNFs, possessing great potential in carrying and stabilizing both water-soluble and oil-soluble natural pigments. In this last section, we further tested if the impact of protein enzymatic hydrolysis on improving HNF formation is universal. An additional set of ten food proteins from common pulses (pea and mung bean), oil seed (peanut), cereals (oat, rice and brown rice), root (potato), algae (chlorella) and milk (whey protein and casein) were evaluated for their ability to form HNFs with copper phosphate.

As can be seen in Supplementary Fig. 9a, most food proteins in their native state formed HNFs with limited quantities and irregular shapes at 0.8 mM $CuSO_4$, where casein failed altogether (a10). It is likely that these food proteins have similarly poor abilities to form HNFs as SPI. Higher $CuSO_4$ concentration at 2.5 mM facilitated the formation of HNFs for all food proteins, but the quantity and homogeneity of all HNFs was still limited. After enzymatic hydrolysis, all food proteins formed HNFs with highly improved quantity and homogeneity at both low (Fig. 5a) and high (Fig. 5b)) metal ion concentrations due to the increased nucleation sites of HNFs and increased flexibility of proteins, especially for whey protein (a9), rice protein (a5) and casein (a10) at 0.8 mM $CuSO_4$ which hardly formed or did not form HNF before enzymatic hydrolysis. Moreover, almost all HNFs displayed smaller sizes at both metal ion concentrations compared to those formed before hydrolysis, leading to potentially higher surface-to-weight ratios. Overall, enzymatic hydrolysis universally shows to be promoting HNF formation in all 11 food proteins, where without this treatment HNF forming capability is limited. Moreover, enzymatic hydrolysis leads to more homogeneous and regular HNFs with potentially increased surface-to-weight ratios needed for successful loading applications.

## Conclusions

We have shown that enzymatic hydrolysis can effectively improve the limited ability of 11 common food proteins to form hybrid nanoflowers (HNFs) by increasing the nucleation sites and enhancing the flexibility of proteins after decomposing them into polypeptides. Soy protein hydrolysate (SPH) at the enzyme-to-substrate ratio of 0.15% displays good performance in combining with various metal ions ($Cu^{2+}$, $Mn^{2+}$, $Fe^{2+}$ and $Ca^{2+}$) and formed various HNFs with different hierarchical structures. The metal phosphates generated by these ions mainly form the scaffolds of HNFs, and SPH primarily acts as the "flesh" to further build up the HNFs. The resulting structures are plump and petal-like. Moreover, a $CuSO_4$-based HNF shows great capability in loading and stabilizing the water-soluble *Monascus* red and oil-soluble curcumin, where the HNF and pigments were combined mainly through hydrogen bonds and hydrophobic interactions, respectively. In conclusion, enzymatic hydrolysis is a powerful tool to fabricate food protein-based HNFs, which have great potential in acting as novel delivery systems for application of sensitive natural pigments in food, cosmetic and medicine.

# Figures

**Fig. 1 Enzymatic hydrolysis by papain improves the ability of soy protein isolate (SPI) to form hybrid nanoflowers (HNFs). a**, Bright field micrographs showing the formation of HNFs at an enzyme-to-substrate (E/S) weight ratio of 0% (a1), 0.015% (a2), 0.06% (a3) and 0.15% (a4). Red arrows indicate the primary precipitates. The scale bars represent 10 μm (inset) and 100 μm. **b**, SDS-PAGE pattern of the protein hydrolysates at different E/S ratios under reducing condition. Lane M: marker, lane 1: 0%, lane 2: 0.015%, lane 3: 0.06%, lane 4: 0.15%. A total of 10 μg protein was loaded in lane 1-4. **c**, The degree of hydrolysis (DH) of SPI at different E/S ratios. **d**, FTIR spectra of

soy protein hydrolysates at different E/S ratios at the amide I region. **e**, Intrinsic emission fluorescence spectra of soy protein hydrolysates at different E/S ratios at 0.1 mg/mL protein. Red arrow indicates the blue shift of the peak.

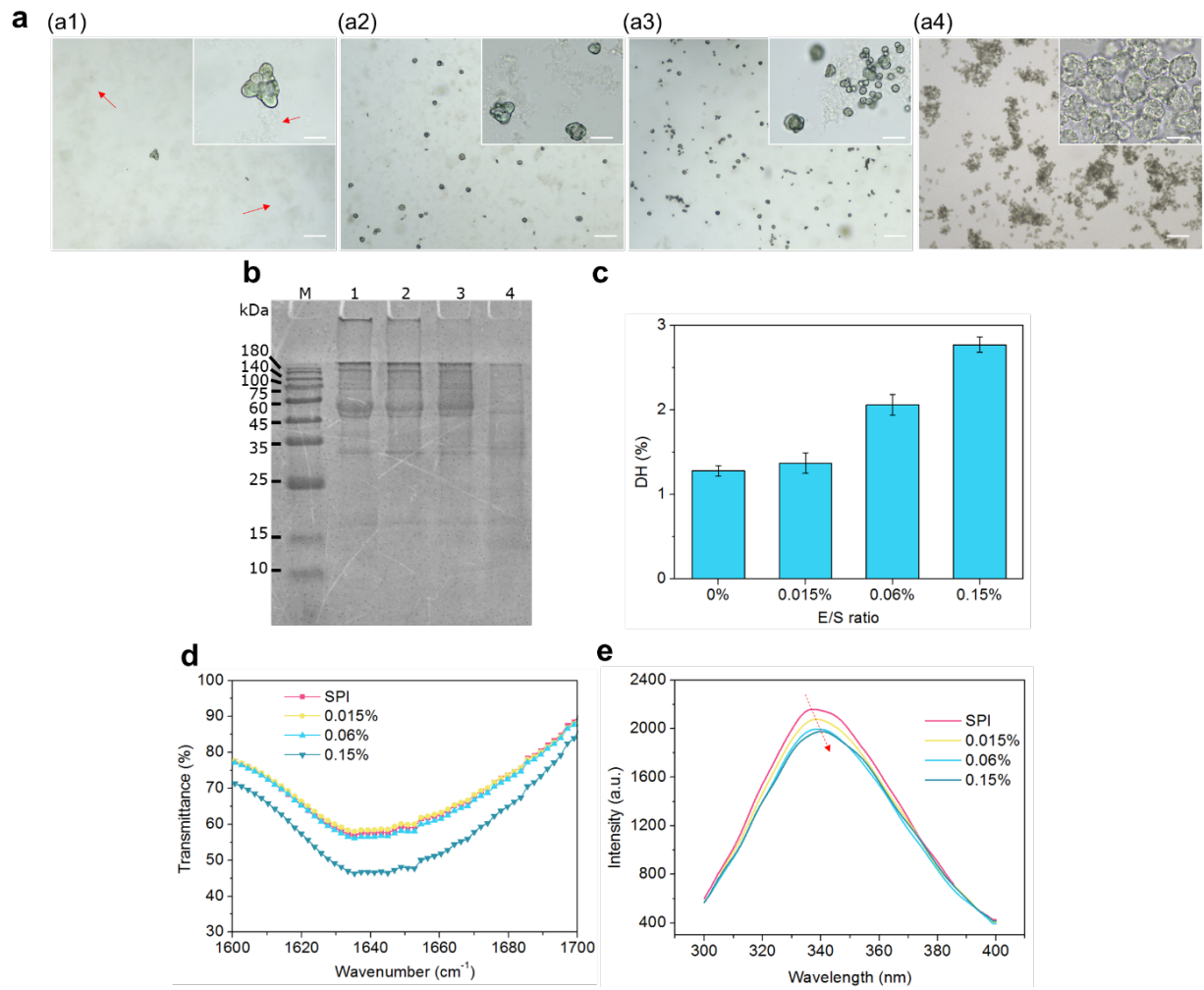

**Fig. 2 Hybrid nanoflowers (HNFs) formation capability of soy protein hydrolysate (SPH) with varied metal ions. a,b**, The morphology of HNFs generated at 0.8 mM (**a**) and 2.5 mM (**b**) metal ion including Cu2+ from CuSO4 (**a1**,**b1**), Mn2+ (**a2**,**b2**), Fe2+ (**a3**,**b3**), Ca2+ (**a4**,**b4**) and Cu2+ from CuCl2 (**a5**,**b5**) by the observation under light microscopy. The insets are macroscopic morphologies of HNFs in the form of precipitates. The scale bars represent 10 μm. **c,d,** The FE-SEM images of HNFs correspond to those in **a**,**b**.

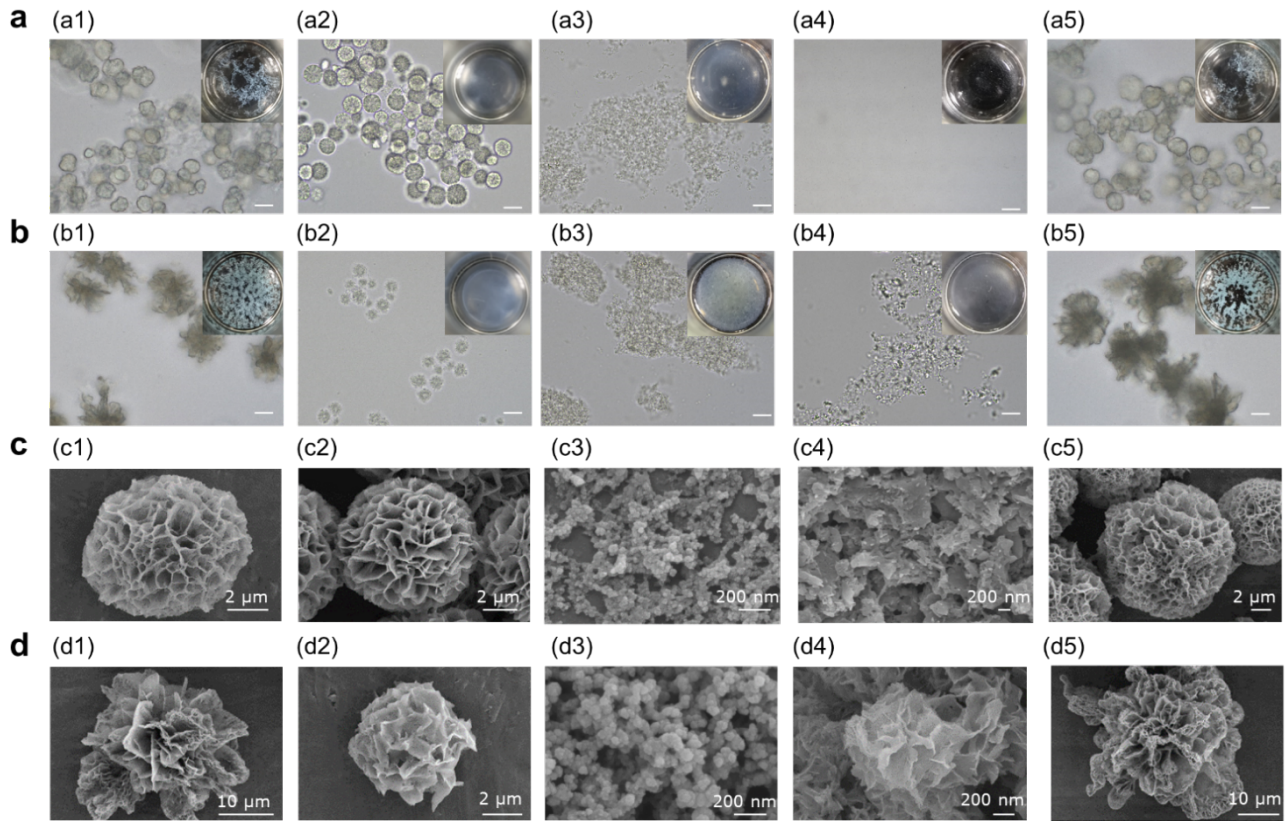

**Fig. 3 Characterisation of various hybrid nanoflowers (HNFs). a,b** Element maps of HNFs prepared at metal ion concentrations of 0.8 mM (**a**) and 2.5 mM (**b**) including Cu2+ from CuSO4 (**a1,b1**), Mn2+ (**a2,b2**), Fe2+ (**a3,b3**), Ca2+ (**b4**) and Cu2+ from CuCl2 (**a5,b5**). **c**, FTIR spectra of HNFs. The dash lines indicate the characteristic absorption peaks of protein, and the arrows refer to the characteristic absorption peaks of $PO_4^{3-}$ group. **d**, XRD patterns of HNFs.

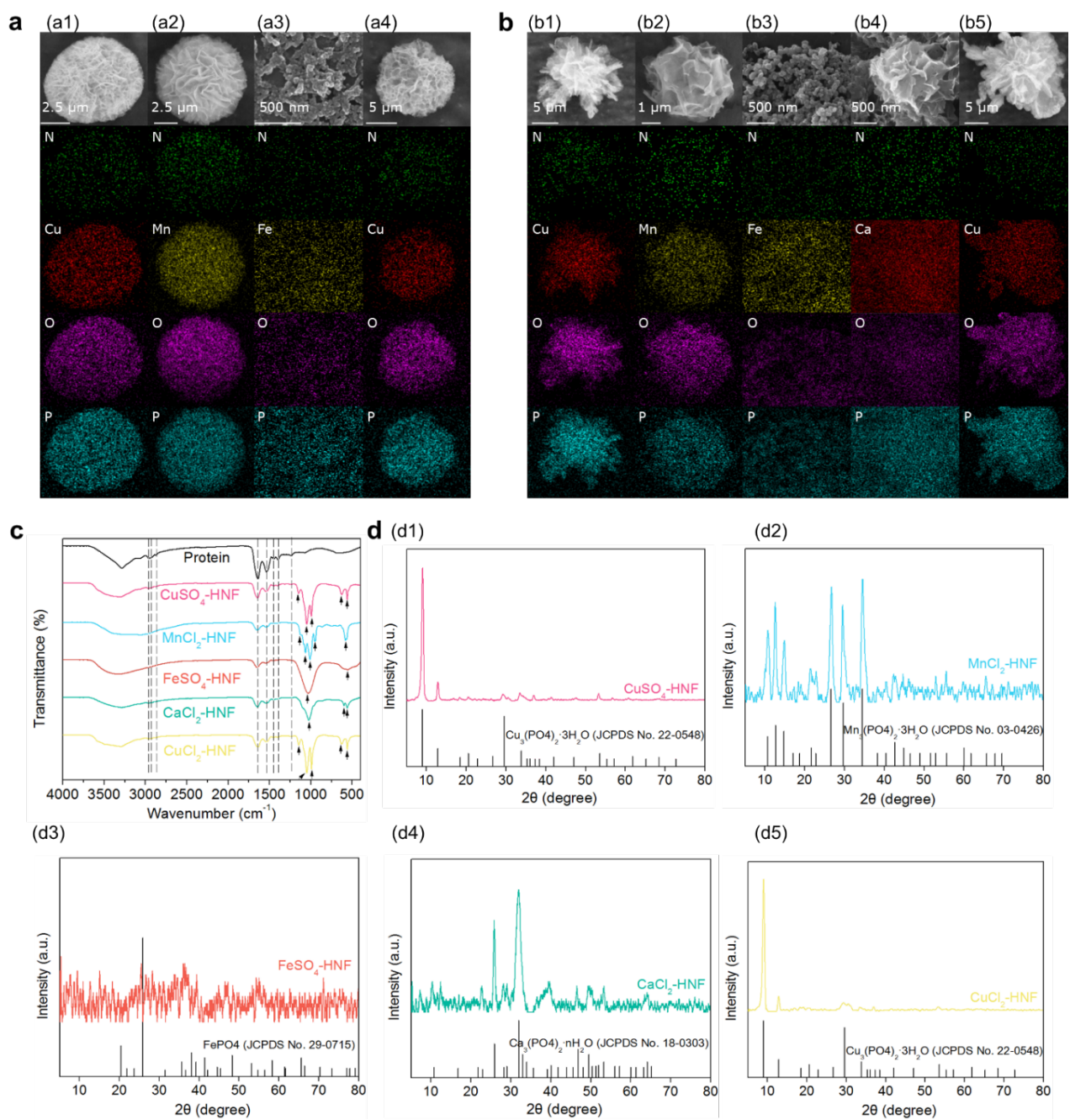

**Fig. 4 Natural pigments loading and stabilization by the hybrid nanoflower prepared with CuSO₄ (CuSO₄-HNF). a**, Macroscopic appearances of curcumin solution and curcumin-HNF complex dispersion before (**a1**,**a2**) and after (**a3**,**a4**) centrifugation, and those of *Monascus* red solution and *Monascus* red-HNF complex dispersion before (**a5**,**a6**) and after (**a7**,**a8**) centrifugation. The solvent is water containing 0.03% xanthan gum to suspend HNFs. **b**, The color change of free curcumin (**b1**) and *Monascus* red (**b4**) and loaded curcumin (**b2**) and *Monascus* red (**b5**), and the corresponding retention rates of these pigments (**b3**,**b6**) under the UV light treatment at 8 W for 2 h. **c**, The color change of free curcumin (**c1**) and *Monascus* red (**c4**) and loaded curcumin (**c2**) and *Monascus* red (**c5**), and the corresponding retention rates of these pigments (**c3**,**c6**) under the heating treatment at 95 °C for 2 h.

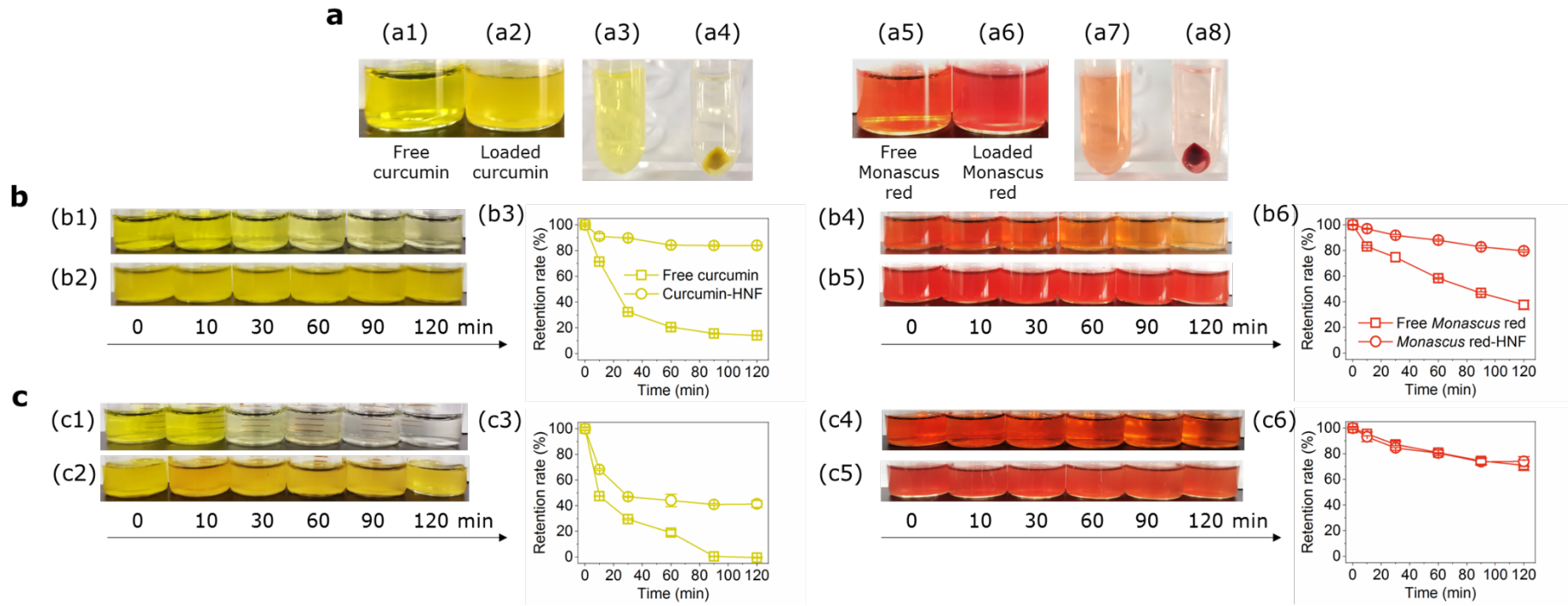

**Fig. 5 The performance of enzymatic hydrolysis in promoting versatile food protein samples to form hybrid nanoflowers (HNFs). a,b,** The morphologies of HNFs formed at 0.8 mM (**a**) and 2.5 mM (**b**) CuSO$_4$ with hydrolysates of pea protein isolate (**a1,b1**), mung bean protein isolate (**a2,b2**), peanut protein isolate (**a3,b3**), oat protein isolate (**a4,b4**), rice protein isolate (**a5,b5**), brown rice protein isolate (**a6,b6**), potato protein isolate (**a7,b7**), chlorella protein concentrate (**a8,b8**), whey protein isolate (**a9,b9**) and casein (**a10,b10**) by the observation under light microscopy. The scale bars represent 10 μm (inset) and 100 μm.

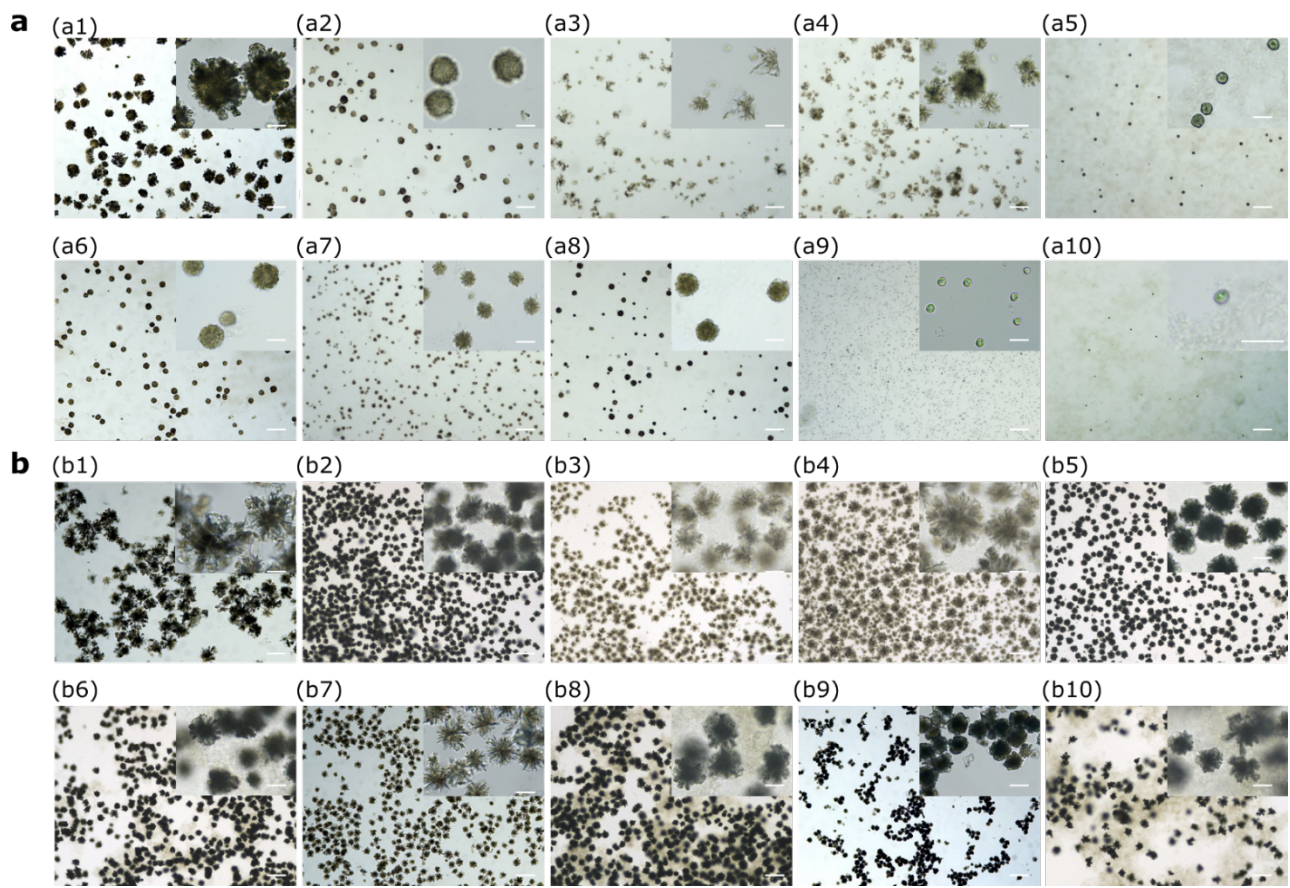

## Methods

**Enzymatic hydrolysis of food proteins.** Each food protein sample was dispersed in water at concentration of 4% (w/w) and stirred for at least 2 h, followed with storing in fridge at 4 °C to ensure complete hydration. Proteins were subjected to hydrolysis by Papain (pH7, 55 °C) at the enzyme-to-substrate (E/S) ratio between 0.015-0.15% (w/w) for 10 min. After the enzymatic hydrolysis, protein hydrolysates were kept in ice bath to inhibit the hydrolysis reaction before further use. All food protein samples contain no less than 80% protein, except for chlorella protein that contains 70% protein.

**Preparation of food protein-based organic-inorganic hybrid nanoflowers (HNFs).** The protein hydrolysates were diluted to protein concentration of 0.1 mg/mL by using phosphate buffer solution (PBS, pH7.4) that contained 10.14 mM Na$_2$HPO$_4$·12H$_2$O, 1.76 mM NaH$_2$PO$_4$·2H$_2$O, 136.89 mM NaCl and 2.68 mM KCl. Afterwards, 16 μL or 50 μL of 150 mM metal compound solution was added into each diluted protein hydrolysate to reach a final metal ion concentration of 0.8 mM or 2.5 mM. The mixtures were fully blended and kept to stand at room temperature for 3 days.

**Characterization of the protein hydrolysates.** Protein composition was determined by sodium dodecyl sulfate-polyacrylamide gel electrophoresis (SDS-PAGE) under the reducing condition according to the method established by Laemmli [44]. Degree of hydrolysis was determined through the reaction of primary amino groups of proteins with o-phthaldialdehyde (OPA), where the OPA method was established by Nielsen et al [45]. Protein secondary structures were assessed by measuring the Fourier-transform infrared (FTIR) spectra of samples in KBr pellets using a Nicolet CCR-1 FTIR instrument in the range between 400 and 4000 cm−1. Protein tertiary structure was evaluated by measuring the intrinsic fluorescence spectra of sample on an F-7000 Fluorescence Spectrophotometer at protein concentration of 0.1 mg/mL. The excitation wavelength was set at 283 nm, and the emission spectra was recorded from 300 to 400 nm at a scan speed of 300 nm/min. Excitation and emission slit widths were set at 5 nm.

**Characterization of HNFs.** Immobilization rate of protein was determined by measuring the protein contents in the supernatants of protein samples and HNF samples based on the method established by Bradford [46]. The supernatants were obtained by centrifuging the samples at 10,000 × g for 10 min. The morphology of HNFs were observed under an Olympus MD-130 optical light microscopy or a Zeiss Merlin field-emission scanning electron microscopy (FE-SEM). The element maps were determined by the energy dispersive X-ray spectroscopy (EDS) equipped to the Zeiss Merlin FE-SEM. The FTIR spectra were obtained on the Nicolet CCR-1 FTIR instrument in the range between 400 and 4000 cm−1. The crystal structures of HNFs were evaluated by X-ray diffraction (XRD) analysis on an Empyrean diffractometer using Mo and Ag radiation. The surface-to-weight ratios of HNFs were determined on a Quantachrome Autosorb iQ Gas Sorption Analyzer.

Pigments loading. A total of 0.15 g CuSO4-HNF was added into 20 mL of pigment solution that contained 400 mg Monascus red in water or 60 mg curcumin in ethanol, and the mixtures were blended for 2 h at 60 °C under magnetic stirring, followed with centrifugation at 10000 × g for 10 min. Pellets were collected and resuspended into 20 mL of water followed by centrifugation. This procedure were repeated for several times to remove free pigments. Finally, the pellets comprised of HNFs with loaded pigments were freeze-dried.

**Characterization of loaded pigments.** Loading amount of pigments in $CuSO_4$-HNF were determined by dissolving 3 mg pigment-HNF complex in 0.1 mL HCl (0.1 M), followed with the addition of 1.4 mL of water for Monascus red-HNF complex or ethanol for curcumin-HNF complex and centrifugation at 3500 × g for 2 min. The supernatants were diluted for 10 folds with water for Monascus red and with ethanol for curcumin, and the absorbance was measured at 486 nm for Monascus red and at 426 nm for curcumin on a YoKe 754N UV–visible spectrophotometer. The pigment concentrations were determined from standard curves (with $R^2$ higher than 0.99) of both free pigments in the same solvents. Interactions between pigments and $CuSO_4$-HNF were assessed by comparing the intrinsic fluorescence spectra and FTIR spectra of the free pigments, HNF and pigment-HNF complexes. For the measurement of the intrinsic fluorescence spectra, $CuSO_4$-HNF and pigment-HNF complexes were suspended in water that contained 0.03% xanthan gum at final pigment concentrations of 0.1 mg/mL for Monascus red and 6 μg/mL for curcumin. The excitation wavelength for Monascus red was set at 450 nm with emission wavelength recorded from 470 to 670 nm, and the excitation wavelength for curcumin was set at 420 nm with emission wavelength recorded from 450 to 670 nm on the F-7000 Fluorescence Spectrophotometer. The FTIR spectra were obtained on the Nicolet CCR-1 FTIR instrument in the range between 400 and 4000 cm$^{-1}$.

**Stability of loaded pigments.** The pigment-HNF complexes were firstly suspended in water that contained

0.03% xanthan gum at final pigment concentrations of 0.1 mg/mL for Monascus red and 6 μg/mL for curcumin, and were subjected to UV light treatment of heating treatment. As for the UV light treatment, the suspensions were poured into petri dishes with liquid height of 4 mm, and the dishes were located at a distance of 10 cm from the UV (8 W) light source for up to 2 h. As regards the heating treatment, the suspensions were heated at 95 °C in water bath for up to 2 h. In both treatments, free pigments in the same solvent were taken as blanks, where the curcumin was previously solubilized in ethanol at concentration of 0.3 mg/mL and was diluted to 6 μg/mL with the water that contained 0.03% xanthan gum. The retention rate of pigment was determined by measuring the pigment contents in both aqueous phase and $CuSO_4$-HNF. The pigment content in aqueous phase was determined from standard curves (with $R^2$ higher than 0.99), and the pigment content in CuSO4-HNF was determined as described aforementioned.

## Acknowledgement

This work was supported by the funding from China Scholarship Council (CSC NO. 202006150032), the National Natural Science Foundation of China (No. 31871746 and 32172157) and Natural Science Foundation of Guangdong Province (No. 2022A1515011598).

## Credit author statement
**Penghui Shen:** Conceptualization, Methodology, Investigation, Validation, Visualization, Writing – original draft. **Mouming Zhao:** Writing - review & editing, Funding acquisition. **Jasper Landman:** Conceptualization, Methodology, Writing – Review & Editing. **Feibai Zhou:** Conceptualization, Methodology, Supervision, Writing – Review & Editing.

## Competing interests
The authors declare no competing interests.

## Supplementary Materials

**Fig. S1** The growth process of the HNF generated with the soy protein hydrolysate at the enzyme-to-substrate (E/S) ratio of 0.15% under 100 × (a-e) and 1000 × (f-j) observation by light microscopy after the mixing of protein dispersion and metal ion solution (0.8 mM) for 0 h (a,f), 3 h (b,g), 12 h (c,h), 24 h (d,i) and 72 h (e,j). The protein dispersion was diluted to 0.1 mg/mL by phosphate buffer (pH 7.4). The scale bars in a-e represent 100 μm, and those in f-j represent 10 μm. Red arrows indicate the primary precipitates of HNFs.

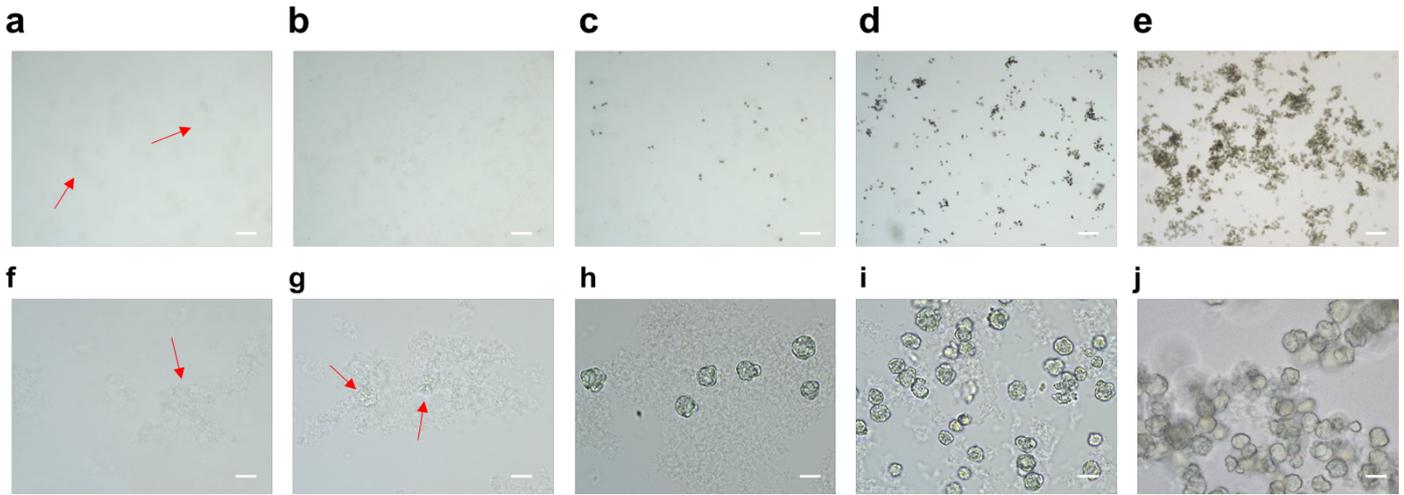

**Fig. S2** The formation of unorganized precipitates at 0.8 mM (a) and 2.5 mM (b) Zn2+ (a1,b1) or Fe3+ (a2,b2). The insets are macroscopic morphologies of the precipitates. The scale bars represent 10 μm.

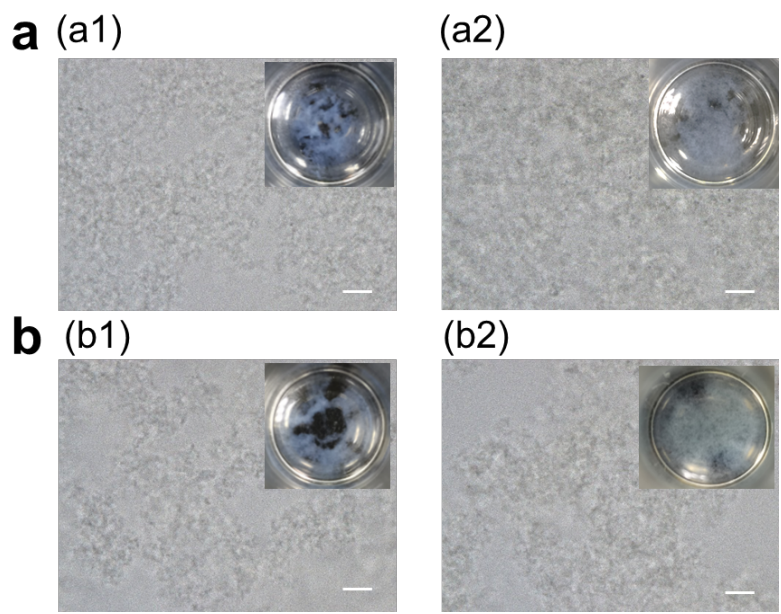

**Fig. S2** The morphology of HNFs generated by soy protein isolate (SPI) at 2.5 mM CuSO4 observed under light microscopy. The inset is the macroscopic morphology of HNFs in the form of precipitates. The scale bar represents 10 μm.

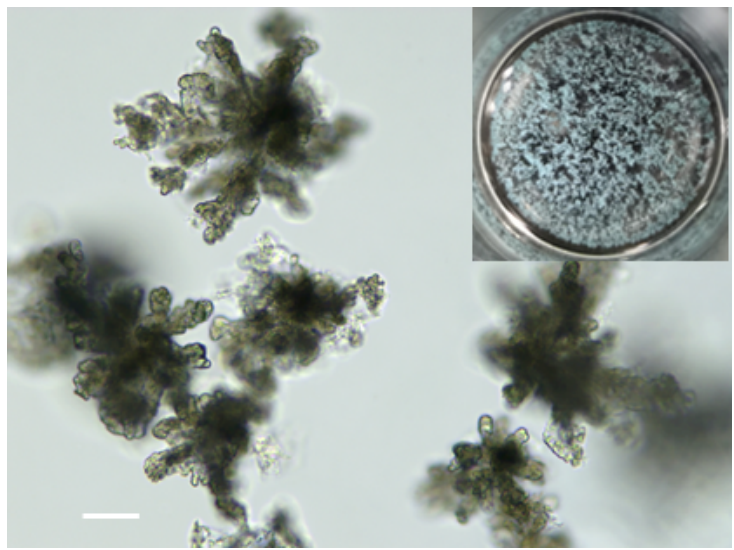

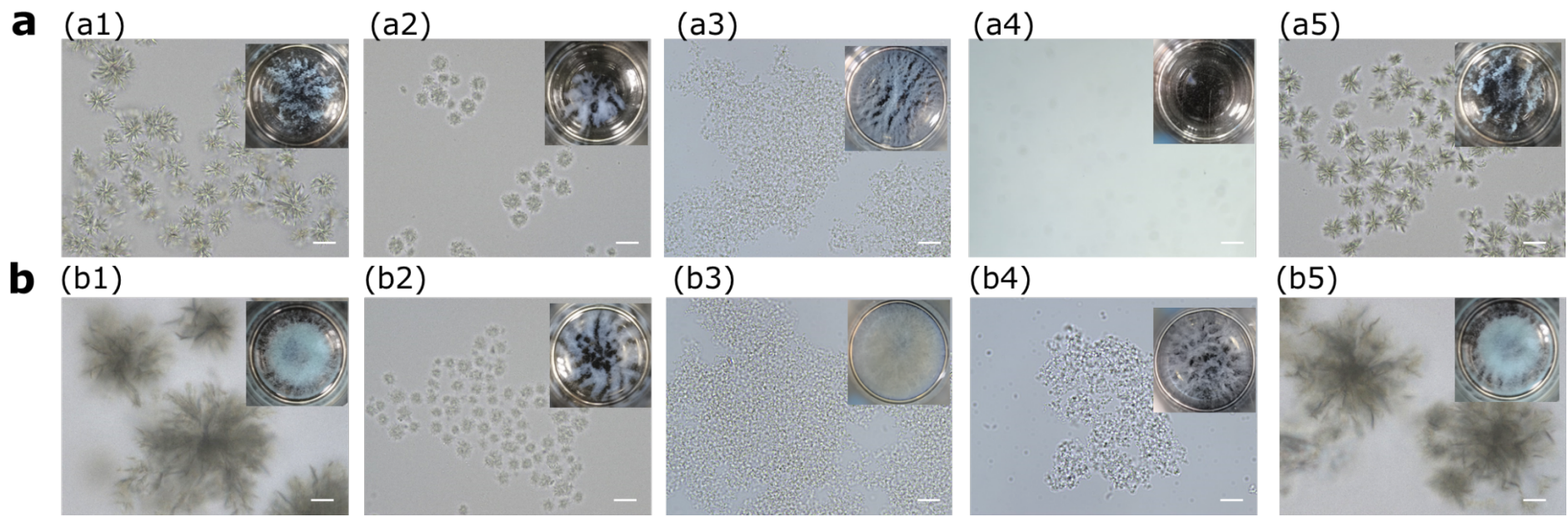

**Fig. S4** The morphology of inorganic nanoflowers generated at 0.8 mM (**a**) and 2.5 mM (**b**) metal ion including $Cu^{2+}$ from $CuSO_4$ (**a1,b1**), $Mn^{2+}$ (**a2,b2**), $Fe^{2+}$ (**a3,b3**), $Ca^{2+}$ (**a4,b4**) and $Cu^{2+}$ from $CuCl_2$ (**a5,b5**) by the observation under light microscopy. The insets are macroscopic morphologies of HNFs in the form of precipitates. The scale bars represent 10 μm.

**Fig. S5** FTIR spectra of curcumin (a) and *Monascus* red (b) before and after loading in CuSO4-HNF.

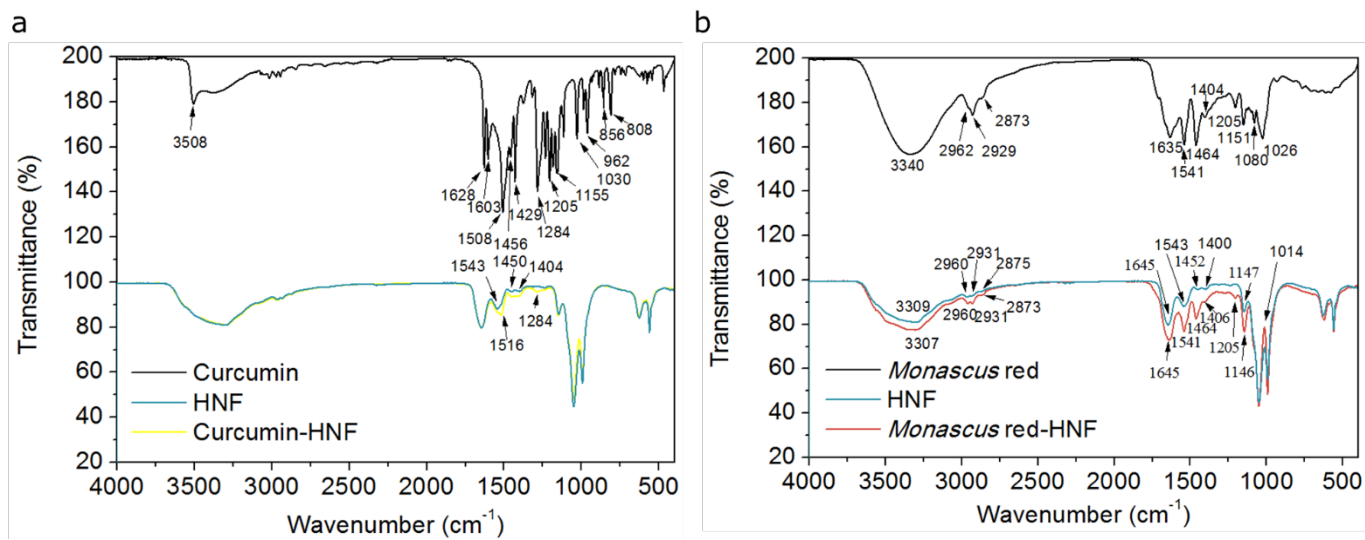

**Fig. S6** The intrinsic emission fluorescence spectra of curcumin (a) and *Monascus* red (b) before and after loading in CuSO4-HNF.

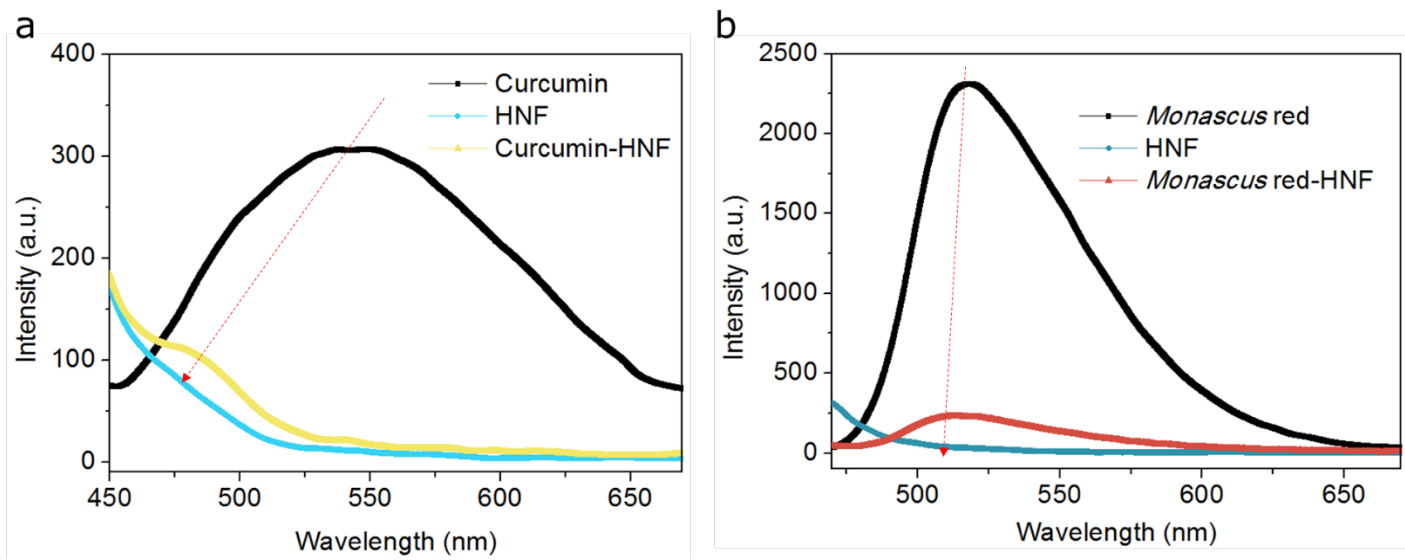

**Fig. S7** The acid-insoluble precipitates generated in curcumin-HNF complex (a) and *Monascus* red-HNF complex (b) during heating treatment at 95 °C.

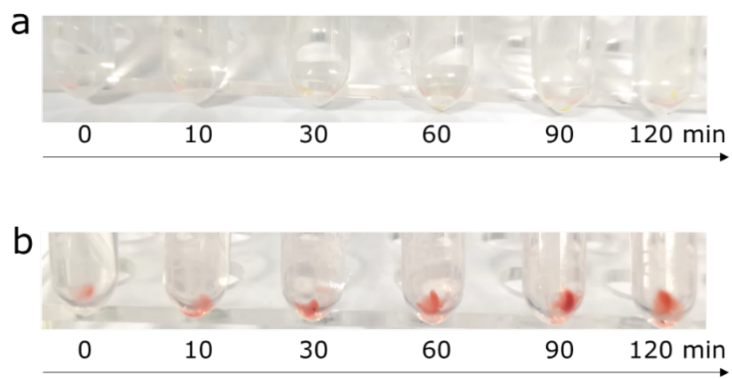

**Fig. S8** The change of retention rate of curcumin (a) and *Monascus* red (b) in CuSO4-HNF and aqueous phase during heating treatment at 95 °C.

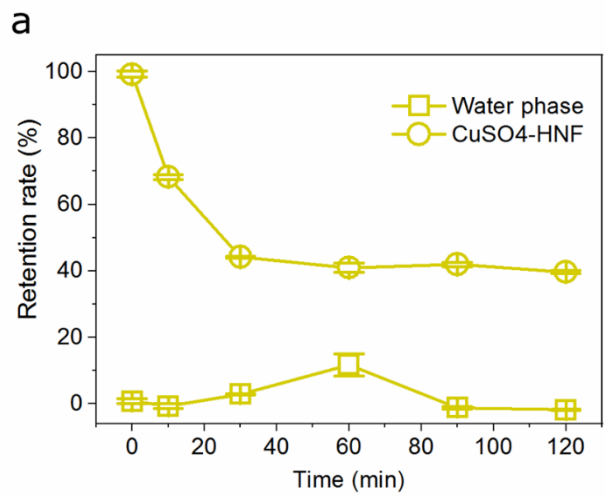 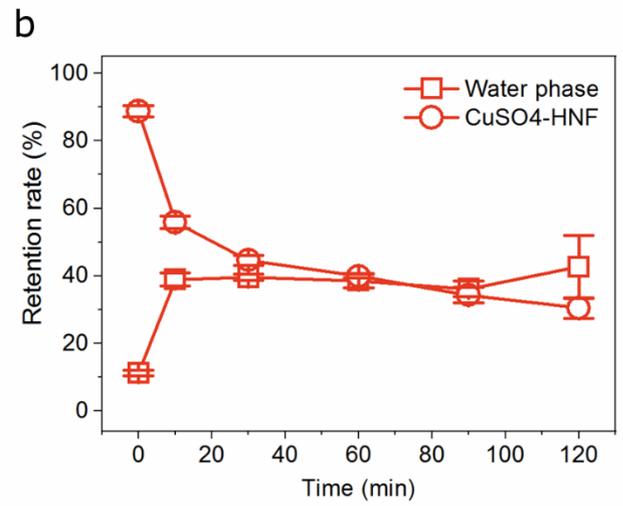

**Fig. S9** The change of retention rate of curcumin (a) and *Monascus* red (b) in CuSO4-HNF and aqueous phase during UV light treatment.

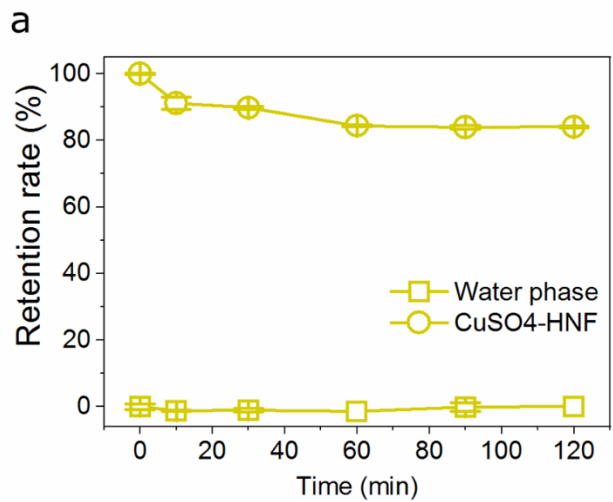
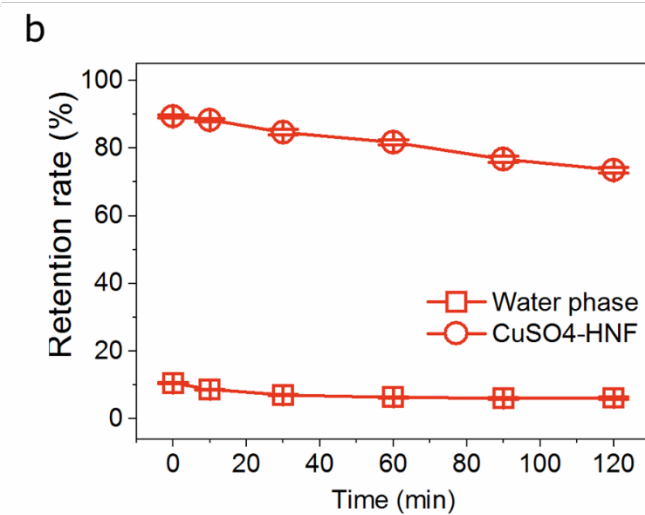

**Fig. S10** a,b, The morphologies of HNFs formed at 0.8 mM (a) and 2.5 mM (b) $CuSO_4$ with pea protein isolate (a1,b1), mung bean protein isolate (a2,b2), peanut protein isolate (a3,b3), oat protein isolate (a4,b4), rice protein isolate (a5,b5), brown rice protein isolate (a6,b6), potato protein isolate (a7,b7), chlorella protein concentrate (a8,b8), whey protein isolate (a9,b9) and casein (a10,b10) by the observation under light microscopy. The scale bars represent 10 µm (inset) and 100 µm.

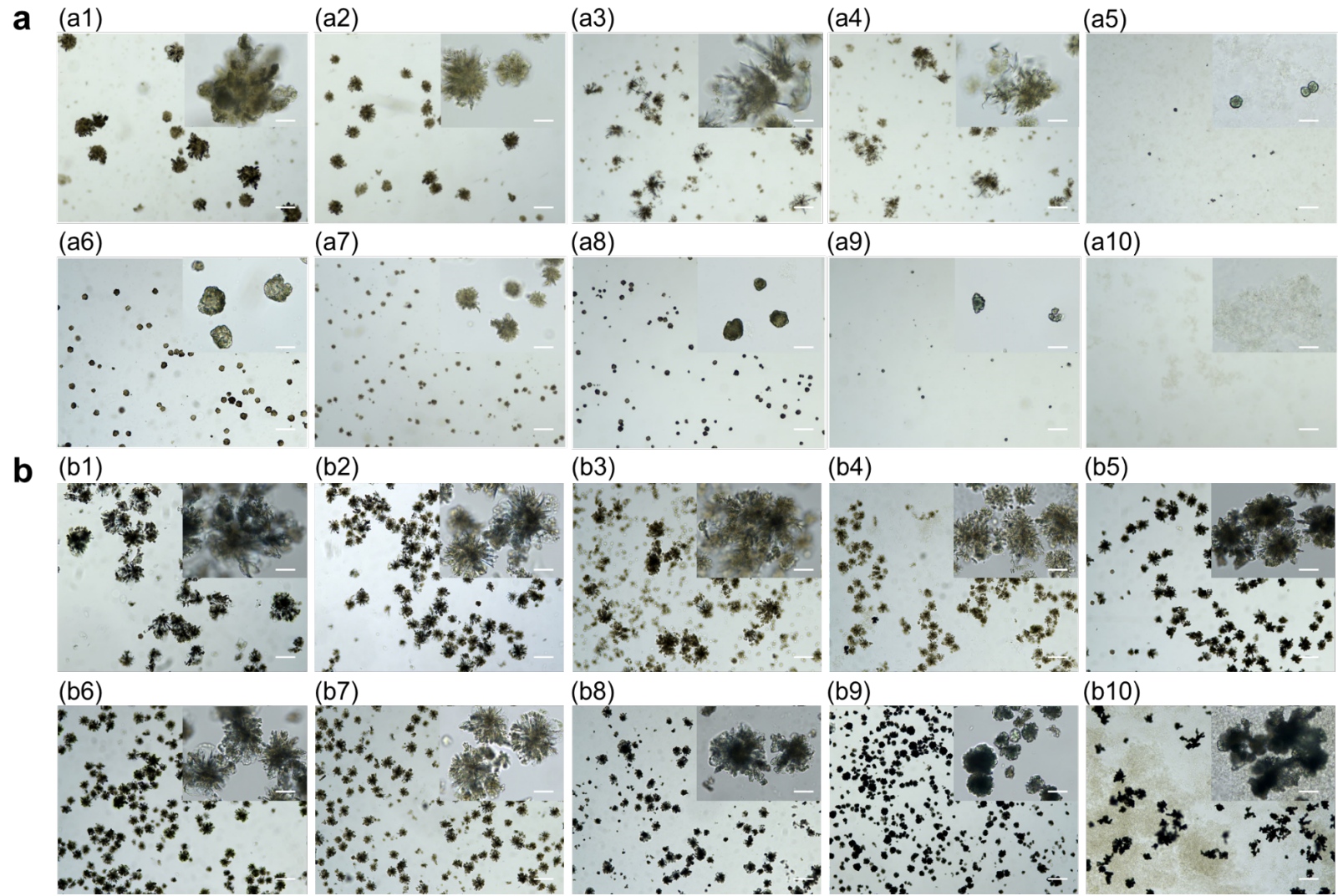

**Table S1** The secondary structure contents of soy protein hydrolysates after the enzymatic hydrolysis by papain.

| Enzyme-to-substrate ratio (%, w/w) | $a$-helix (%) | $\beta$-sheet (%) | $\beta$-turn (%) | Random coil (%) |
|---|---|---|---|---|
| 0 | 18.91 | 32.12 | 24.07 | 24.91 |
| 0.015 | 19.65 | 32.01 | 24.32 | 24.02 |
| 0.06 | 19.68 | 31.54 | 24.33 | 24.44 |
| 0.15 | 20.27 | 30.74 | 25.19 | 23.80 |

**Table S2** The peak and maximum intensity of the intrinsic emission fluorescence spectra of the soy protein hydrolysates at 0.1 mg/mL protein.

| Enzyme-to-substrate ratio (%, w/w) | Peak (nm) | Maximum intensity (a.u.) |
| --- | --- | --- |
| 0 | 337.60 ± 0.42 | 2112.15 ± 66.40 |
| 0.015 | 338.58 ± 0.18 | 2109.85 ± 45.89 |
| 0.06 | 339.73 ± 0.95 | 2024.25 ± 39.24 |
| 0.15 | 340.05 ± 0.28 | 2017.91 ± 57.72 |